\documentclass[conference,romanappendices,10pt]{IEEEtran}
\IEEEoverridecommandlockouts

\usepackage{algorithmic,amsmath,amssymb,amsthm,array,bbm,bibentry,cite,color,comment}
\usepackage{enumerate,enumitem,eurosym,float,graphicx,lettrine,mathrsfs,multirow,nomencl}
\usepackage{pict2e,psfrag,ragged2e,relsize,setspace,subfigure,supertabular,tabularx,url}
\usepackage[english]{babel}

{		\newtheorem{theorem}{Theorem}}
{ 		}
{ 		\newtheorem{lemma}{Lemma}}
{ 		}
{ 		}
{		}
{		}
{ 		\newtheorem{corollary}{Corollary}}

\newcommand{\setL}{\mathcal{L}}

\newcommand{\Real}{\mbox{$\mathbb{R}$}}

\newcommand{\diff}{\mathrm{d}}
\newcommand{\Exp}{\mathbb{E}}

\renewcommand{\Pr}{\mathbb{P}}

\newcommand{\sinr}{\mathrm{SINR}}
\newcommand{\sir}{\mathrm{SIR}}
\newcommand{\los}{\textnormal{\tiny{LOS}}}
\newcommand{\nlos}{\textnormal{\tiny{NLOS}}}

\title{Impact of LOS/NLOS Propagation and Path Loss \\ in Ultra-Dense Cellular Networks}
\author{\IEEEauthorblockN{Jes\'{u}s Arnau, Italo Atzeni, and Marios Kountouris}
\IEEEauthorblockA{Mathematical and Algorithmic Sciences Lab \\ France Research Center, Huawei Technologies Co. Ltd. \\
20 Quai du Point du Jour, 92100 Boulogne-Billancourt, France. \\
Email: \{jesus.arnau, italo.atzeni, marios.kountouris\}@huawei.com}}
\makeindex

\begin{document}

\maketitle

\begin{abstract}
Most prior work on performance analysis of ultra-dense cellular networks (UDNs) has considered standard power-law path loss models and non-line-of-sight (NLOS) propagation modeled by Rayleigh fading. The effect of line-of-sight (LOS) on coverage and throughput and its implication on network densification are still not fully understood. In this paper, we investigate the performance of UDNs when the signal propagation includes both LOS and NLOS components. Using a stochastic geometry based cellular network model, we derive expressions for the coverage probability, as well as tight approximations and upper bounds for both closest and strongest base station (BS) association. Our results show that under standard singular path loss model, LOS propagation increases the coverage, especially with nearest BS association. On the contrary, using dual slope path loss, LOS propagation is beneficial with closest BS association and detrimental for strongest BS association.
\end{abstract}


\section{Introduction} \label{sec:intro}

Network densification is foreseen as one of the key enablers to realize the vision of emerging 5th generation (5G) wireless networks \cite{Bhu14,Que13}. Heterogeneous cellular network (HetNet) deployment is a promising and effective way to provide high cellular network capacity by overlaying conventional macrocell architecture with heterogeneous architectural features, such as small cellular access points (picocells and femtocells), low-power fixed relays, and distributed antennas. Ultra-dense networks (UDNs), i.e., dense and massive deployment of small cells, are expected to achieve higher data rates and enhanced coverage by exploiting spatial reuse, while retaining at the same time seamless connectivity and low energy consumption. Supported by recent studies, UDNs are expected to achieve unprecedented data rates and power consumption reduction. Nevertheless, most prior performance analyses use spatial models, in which the base stations (BSs) are located according to a homogeneous Poisson point process (PPP) and propagation is modeled using standard power-law path loss and Rayleigh fading. For instance, using the aforementioned model with nearest BS association \cite{And11} and strongest BS association \cite{Dhi12}, throughput is shown to grow linearly with the density of BSs per area in the absence of background noise.

Recently, there has been a growing interest in devising increasingly realistic models for system-level performance evaluation of cellular networks. In this respect, \cite{Zha15} studies the impact of dual slope path loss on the performance of downlink UDNs and shows that both coverage and capacity strongly depend on the network density. In \cite{Bai14}, a stochastic geometry based framework for millimeter wave (mmWave) and path loss with line-of-sight (LOS) and non-line-of-sight (NLOS) propagation is proposed. Further models and studies on the effect of LOS propagation in higher frequencies can be found in \cite{Gal15,Din15}. However, all previous studies capture the effect of LOS on the large-scale fading (i.e., path loss), yet always assuming Rayleigh fast fading. This is a coarse simplification, mainly due to tractability, which can significantly alter coverage and throughput performance, since LOS propagation is known to be subject to Ricean fast fading.

In this work, we broaden prior studies on network densification and propose a general stochastic geometry based framework for the effect of LOS/NLOS propagation in both small-scale and large-scale fading. This allows for a practically relevant analysis of UDNs under a more realistic setting, which models LOS propagation using Ricean fading; more precisely, we approximate the Rician fading by Nakagami-$m$ distribution for tractability. Remarkably, the proposed framework accommodates generalized distance-dependent LOS probability functions; in addition, it encompasses both closest and strongest BS association, as well as single- and multi-slope path loss. As a particular scenario, we consider the ITU-R urban micro-cell (UMi) LOS probability model \cite{UMi} and provide a tractable approximation for the coverage probability. Our results provide crisp insights into the coverage and throughput performance and show the impact of LOS on BS association. We show that, under standard power-law path loss model, LOS propagation increases the coverage, especially with closest BS association. On the contrary, considering dual- or multi-slope path loss models, LOS propagation is beneficial for closest BS association and detrimental for strongest BS association.

\section{System model} \label{sec:sys_model}

We consider a typical downlink user equipment (UE) located at the origin of the Euclidean plane. For the location distribution of the BSs, we consider the marked Poisson point process (PPP) $\widehat{\Phi} \triangleq \{ (x_{i}, h_{x_{i}}) \} \subset \Real^{2} \times \Real^{+}$, where the underlying point process $\Phi \triangleq \{ x_{i} \} \subset \Real^{2}$ is a homogeneous PPP with density $\lambda$ and the mark $h_{x_{i}} \in \Real^{+}$ represents the channel power fading gain from the BS located at $x_{i}$ to the typical UE. Let $\ell:\Real^{+} \to \Real^{+}$ denote the path loss function. In the first part or unless otherwise stated, we assume the standard power-law path loss model with $\ell(r_{x_{i}}) = r_{x_{i}}^{-\alpha}$, where $r_{x_{i}} \triangleq \| x_{i} \|$. In Section~\ref{sec:multislope}, we generalize our results for a multi-slope path loss model. Lastly, we assume that all BSs transmit with unit power. 

The signal-to-interference-plus-noise ratio (SINR) when the typical UE is associated to the BS located at $x$ is given by
\begin{equation} \label{eq:SINR}
\sinr_{x} \triangleq \frac{h_{x} \ell(r_{x})}{I + \sigma^{2}}
\end{equation}
where $I$ is the overall interference term defined as
\begin{equation}
I \triangleq \sum\limits_{y \in \Phi \backslash \{ x \}} h_{y} \ell(r_{y})
\end{equation}
and $\sigma^{2}$ is the additive noise power.

In our model, each BS is characterized by either LOS or NLOS propagation independently from the others and regardless of its role as serving or interfering BS. 
Assuming a distance-dependent LOS probability function $p_{\los}(r_{x})$, i.e., the probability that a BS located at $x$ experiences LOS propagation, which depends on the distance $r_{x}$, the distribution of the channel power gain is expressed as
\begin{equation}
f_{h_{x}}(z) \triangleq p_{\los}(r_{x}) f_{\los}(z) + (1 - p_{\los}(r_{x})) f_{\nlos}(z)
\end{equation}
where $f_{\los}(z)$ and $f_{\nlos}(z)$ are the pdfs corresponding to LOS and NLOS propagation, respectively. In the following, we assume that the channel amplitudes are Rayleigh distributed in NLOS propagation condition and Nakagami-$m$ distributed in LOS conditions; hence, the channel power gains are distributed according to exponential and Gamma distributions, respectively. The complementary cdf of the latter is given by
\begin{equation} \label{eq:gamma_ccdf}
\bar{F}_{\los}(z) \triangleq 1-\frac{\gamma(m, m z)}{\Gamma(m)} = \exp(-m z) \sum_{k=0}^{m-1} \frac{(m z)^k}{k!}
\end{equation}where the last equality holds when $m$ is an integer; note that $\bar{F}_{\nlos}(z)$ can be obtained from \eqref{eq:gamma_ccdf} simply by setting $m=1$. Approximating Ricean fading with a Nakagami-$m$ distributed amplitude is common practice because of its tractability, flexibility, and good fitting performance. In the general LOS case, $m$ is computed as $m \triangleq (K+1)^{2}/(2K+1)$, where $K$ is the Ricean $K$-factor representing the ratio between the powers of the direct and scattered paths.\footnote{Note that in order to use the simplest formulation in \eqref{eq:gamma_ccdf}, the value of $m$ is rounded to the closest integer.} For simplicity, we focus on the interference-limited case, where $I \gg \sigma^{2}$ in \eqref{eq:SINR} and hence work with the signal-to-interference ratio (SIR).

\section{SIR coverage with LOS propagation} \label{sec:coverage_probability}

In this section, we provide our most general result, which is an expression for the coverage probability when both serving and interfering BSs independently experience either LOS or NLOS conditions, depending on their distance from the typical UE; note that we define the coverage probability as the probability that the received SIR is larger than a target $\theta$, i.e., $\mathrm{P}_{\mathrm{cov}}(\theta) = \mathbb{P}(\sir>\theta)$; both closest \cite{And11} and strongest (i.e., highest SINR) \cite{Dhi12} BS association are considered.

The result is shown below in Theorem~\ref{th:main}. Before, let us introduce the following preliminary definitions. We define
\begin{align}
\label{eq:phi} \phi(r) & \triangleq \left\{
\begin{array}{ll}
2 \pi \lambda e^{- \pi \lambda r^{2}} r, & \quad \textrm{closest BS} \\
2 \pi \lambda r, & \quad \textrm{strongest BS}
\end{array} \right. \\
\label{eq:nu} \nu(r) & \triangleq \left\{
\begin{array}{ll}
r, & \hspace{16mm} \quad \textrm{closest BS} \\
0, & \hspace{16mm} \quad \textrm{strongest BS}
\end{array} \right.
\end{align}
which allow us to generalize our results for both closest and strongest BS association. Furthermore, we use $\mathrm{P}_{\mathrm{cov}}^{\mathrm{\nlos}}(\theta)$ and $\setL_{I}^{\nlos}(s)$ to denote the coverage probability and the Laplace transform of the interference, respectively, when there is only NLOS propagation:
\begin{align}
\label{eq:P_cov_nlos} \mathrm{P}_{\mathrm{cov}}^{\mathrm{\nlos}}(\theta) & \triangleq \int_{0}^{\infty} \setL_{I}^{\nlos}(\theta r^{\alpha}) \phi(r) \, \diff r \\
\label{eq:LI_nlos} \setL_{I}^{\nlos}(s) & \triangleq \exp \bigg( -2 \pi \lambda \int_{\nu(r)}^{\infty} \bigg( 1 - \frac{1}{1 + s t^{-\alpha}} \bigg) t \,\diff t \bigg).
\end{align}

\begin{theorem} \label{th:main} \rm{
The coverage probability is given by \eqref{eq:pcov_general} at the top of the next page, where $\setL_{I}^{\los}(s)$ is the Laplace transform of the interference when BSs experience LOS or NLOS conditions:
\begin{align}
\label{eq:LI_main} \setL_{I}^{\los}(s) & \triangleq \setL_{I}^{\nlos}(s) \\
\nonumber & \hspace{-11.5mm} \times \exp \! \bigg( \! \!  - 2 \pi \lambda \int_{\nu(r)}^{\infty} \! \! \! \! \! \! p_{\los}(t) \bigg( \frac{1}{1 + s t^{-\alpha}} - \frac{1}{(1 + \frac{s}{m} t^{-\alpha} )^{m}} \bigg) t \, \diff t \bigg).
\end{align}}
\end{theorem}

\begin{figure*}[t!]
\begin{align}
\label{eq:pcov_general} \mathrm{P}_{\mathrm{cov}}(\theta) & \triangleq \mathrm{P}_{\mathrm{cov}}^{\mathrm{\nlos}}(\theta) + \mathlarger{{\int_{0}^{\infty}}} p_{\los}(r) \bigg( \bigg[ \sum_{k=0}^{m - 1} \frac{(-s)^{k}}{k!} \frac{\diff^{k}}{\diff s^{k}} \setL_{I}^{\los}(s) \bigg]_{s = m \theta r^{\alpha}} \! \! \! \! - \setL_{I}^{\nlos} (\theta r^{\alpha}) \bigg) \phi(r) \, \diff r
\end{align}
\hrulefill
\end{figure*}

\begin{IEEEproof}
See Appendix~\ref{sec:app_main}.
\end{IEEEproof} \vspace{1mm}

The expressions above are valid for any integrable $p_{\los}(r)$. However, the particular characteristics of this function will determine how easily the integrals can be evaluated. Inspired by 3GPP channel models, in the following section, we propose a model for $p_{\los}(r)$ that greatly simplifies the resulting expressions, while still capturing the underlying physical phenomena.

\section{Tractable LOS Probability Model} \label{sec:pLOS}
Several LOS probability models have been proposed in the literature, most of them distance-dependent. A commonly used model is the ITU-R UMi model \cite {UMi} (referred to as 3GPP model in the following), given by
\begin{equation} \label{eq:3GPP}
p_{\los}(r) = \min \bigg( \frac{18}{r}, 1 \bigg) \big( 1 - e^{-\tfrac{r}{36}} \big) + e^{-\tfrac{r}{36}}
\end{equation} 
where the propagation is always in LOS conditions for $r \leq 18$~m. In practice, this implies that, for densities above $\lambda=10^{-2}$~BS/m$^{2}$ and closest BS association, the probability of LOS coverage is very close to one. As a consequence, some NLOS terms in the previous expressions could be neglected.

Following this line of thought, we propose the following simplified model that preserves the flatness of the first part of the curve and can be used for analytical calculations; our numerical results will show that it approximates very accurately widely used 3GPP channel models. Its expression is given by
\begin{align} \label{eq:3GPP_simpl}
p_{\los}(r) = \left\{
\begin{array}{ll}
1, & r \in (0, D] \\
0, & r \in (D, \infty)
\end{array}
\right.
\end{align}
with $D>0$ being the critical distance below which all BSs are in LOS conditions. In this scenario, the coverage probability in \eqref{eq:pcov_general} simplifies as
\begin{align}
\label{eq:Pcov_closest_simpl} \mathrm{P}_{\mathrm{cov}}(\theta) = \ & \int_{D}^{\infty} \setL_{I}^{\nlos} (\theta r^{\alpha}) \phi(r) \, \diff r \\
\nonumber & + \int_{0}^{D} \bigg[ \sum_{k=0}^{m - 1} \frac{(-s)^{k}}{k!} \frac{\diff^{k}}{\diff s^{k}} \widetilde{\setL_{I}^{\los}}(s) \bigg]_{s = m \theta r^{\alpha}} \!\!\!\phi(r) \, \diff r.
\end{align}
Here, $\setL_{I}^{\nlos}(s)$ is the Laplace transform of the interference when all the BS are in NLOS conditions introduced in \eqref{eq:LI_nlos}. On the other hand, the Laplace transform comprising both LOS and NLOS conditions admits now a much simpler expression, denoted above by $\widetilde{\setL_{I}^{\los}}(s)$ and given by
\begin{align}
\nonumber \widetilde{\setL_{I}^{\los}}(s) & \triangleq \exp \bigg( -2 \pi \lambda \int_{\nu(r)}^{D} \bigg( 1 - \frac{1}{1 + s t^{-\alpha}} \bigg) t \; \diff t \bigg) \\
& \hspace{-5mm} \times \exp \bigg( -2 \pi \lambda \int_{\nu(r)}^{D} \bigg( 1 - \frac{1}{\big( 1 + \frac{s}{m} t^{-\alpha} \big)^{m}} \bigg) t \; \diff t \bigg).
\end{align}

\begin{corollary} \label{cor:lim} \rm{
The coverage probability in \eqref{eq:Pcov_closest_simpl} when $\lambda \to 0$ becomes 
\begin{equation} \label{eq:lim}
\lim_{\lambda \to 0} \mathrm{P}_{\mathrm{cov}}(\theta) = \mathrm{P}_{\mathrm{cov}}^{\nlos}(\theta)
\end{equation}
where $\mathrm{P}_{\mathrm{cov}}^{\nlos}(\theta)$ is the coverage probability when there is only NLOS propagation defined in \eqref{eq:P_cov_nlos}.}
\end{corollary}

\begin{IEEEproof}
See Appendix~\ref{sec:app_lim}.
\end{IEEEproof} \vspace{1mm}

These expressions can now be evaluated by resorting to numerical integration and differentiation. However, the latter can be cumbersome in practice, especially for large values of $m$. Thus, to make numerical evaluation more efficient, a simpler tractable upper bound with no derivatives is sometimes more practical. The following Lemma provides a result that can be used for this purpose.
\begin{lemma} \label{lem:closest_simpl} \rm{
For the summation in (\ref{eq:Pcov_closest_simpl}), the following inequality holds:
\begin{align}\label{eq:bounds}
\nonumber & \hspace{-2mm} \bigg[ \sum_{k=0}^{m-1} \frac{(-s)^{k}}{k!} \frac{\diff^{k}}{\diff s^{k}} \setL_{I}(s) \bigg]_{s = m \theta r^{\alpha}} \\
& \leq \sum_{k=1}^{m} (-1)^{k+1} \binom{m}{k} \setL_{I} \Big( (\Gamma(m+1))^{-\tfrac{1}{m}} k m \theta r^{\alpha} \Big).
\end{align}}
\end{lemma}

\begin{IEEEproof}
See Appendix~\ref{sec:app_closest_simpl}.
\end{IEEEproof}

\section{Impact of Multi-Slope Path loss on NLOS/LOS Coverage} \label{sec:multislope}

So far we have assumed a simple power-law path loss model with a single path loss exponent. However, the general framework provided in Section~\ref{sec:coverage_probability} admits a straightforward extension to multi-slope path loss models with $N$ different path loss exponents $\{ \alpha_{n} \}_{n=0}^{N-1}$ (cf. \cite{Zha15}). In particular, the coverage probability is given by \eqref{eq:Pcov_closest_dual} at the top of the next page, where the values $\{R_{n}\}$ mark the transition distances between the different path loss exponents $\{\alpha_{n}\}$; we note that, for coherence, $R_{0} = 0$ and $R_{N} = \infty$. In \eqref{eq:Pcov_closest_dual}, the Laplace transforms of the interference are given by \addtocounter{equation}{+1}
\begin{equation} \label{eq:LI_nlos_dual}
\setL_{I,N}^{\nlos}(s) \triangleq \prod_{l=0}^{N-1} \exp \bigg( -2 \pi \lambda \int_{\widetilde{R}_{l}}^{\widetilde{R}_{l+1}} \!\! \bigg( 1 - \frac{1}{1 + s t^{-\alpha_{l}}} \bigg) t \,\diff t \bigg)
\end{equation}
and \eqref{eq:LI_main_dual} at the top of the next page: here, we have defined an alternative set of transition distances $\{\widetilde{R}_{n}\}$ that are equal to $\{R_{n}\}$ with the exception of $\widetilde{R}_{0} = \nu(r)$.

\begin{figure*}
\begin{align} \addtocounter{equation}{-2}
\label{eq:Pcov_closest_dual} \mathrm{P}_{\mathrm{cov},N}(\theta) & \triangleq  \sum_{n=0}^{N-1} \bigg( \int_{R_{n}}^{R_{n+1}} \!\!\! \setL_{I, {N}}^{\nlos} (\theta r^{\alpha_{n}})\phi(r) \, \diff r + \int_{R_{n}}^{R_{n+1}} \!\!\! p_{\los}(r) \bigg( \bigg[ \sum_{k=0}^{m - 1} \frac{(-s)^{k}}{k!} \frac{\diff^{k}}{\diff s^{k}} \setL_{I,N}^{\los}(s) \bigg]_{s = m \theta r^{\alpha_{n}}} \!\!\!\!\!\! - \setL_{I,N}^{\nlos} (\theta r^{\alpha_{n}}) \bigg) \phi(r) \, \diff r \bigg) \\ \addtocounter{equation}{+1}
\label{eq:LI_main_dual} \setL_{I,N}^{\los}(s) & \triangleq \setL_{I,N}^{\nlos}(s) \prod_{l=0}^{N-1} \exp \bigg( - 2 \pi \lambda  \int_{\widetilde R_{l}}^{\widetilde R_{l+1}}\!\! p_{\los}(t) \bigg( \frac{1}{1 + s t^{-\alpha_{l}}} - \frac{1}{(1 + \frac{s}{m} t^{-\alpha_{l}})^{m}} \bigg) t \, \diff t \bigg)
\end{align}
\hrulefill \vspace{-2mm}
\end{figure*}

\begin{figure*}[t!]
	\centering
	\includegraphics[scale=1]{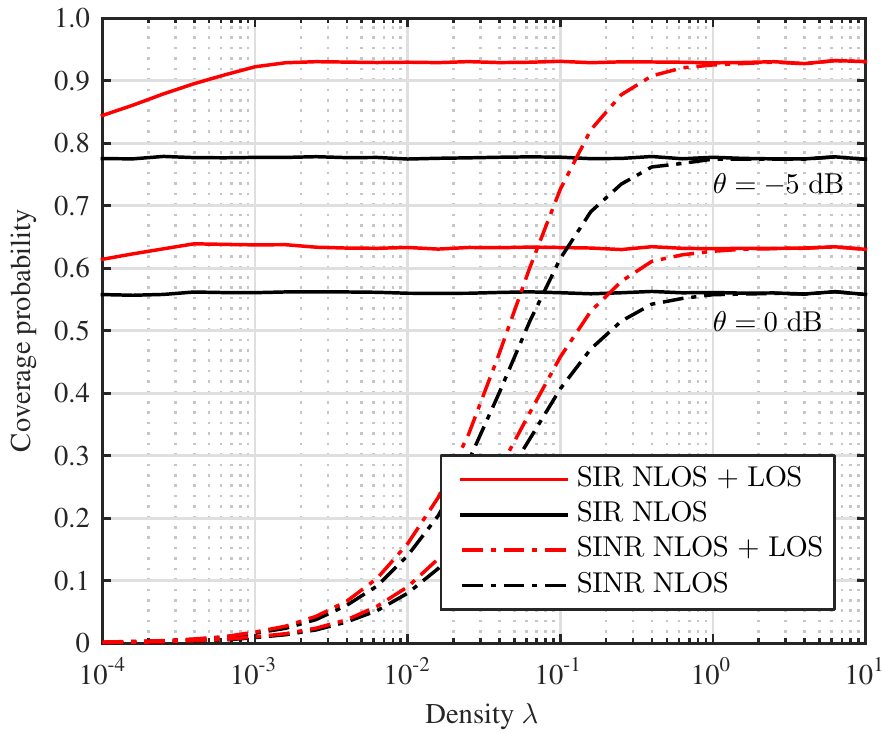}
	\includegraphics[scale=1]{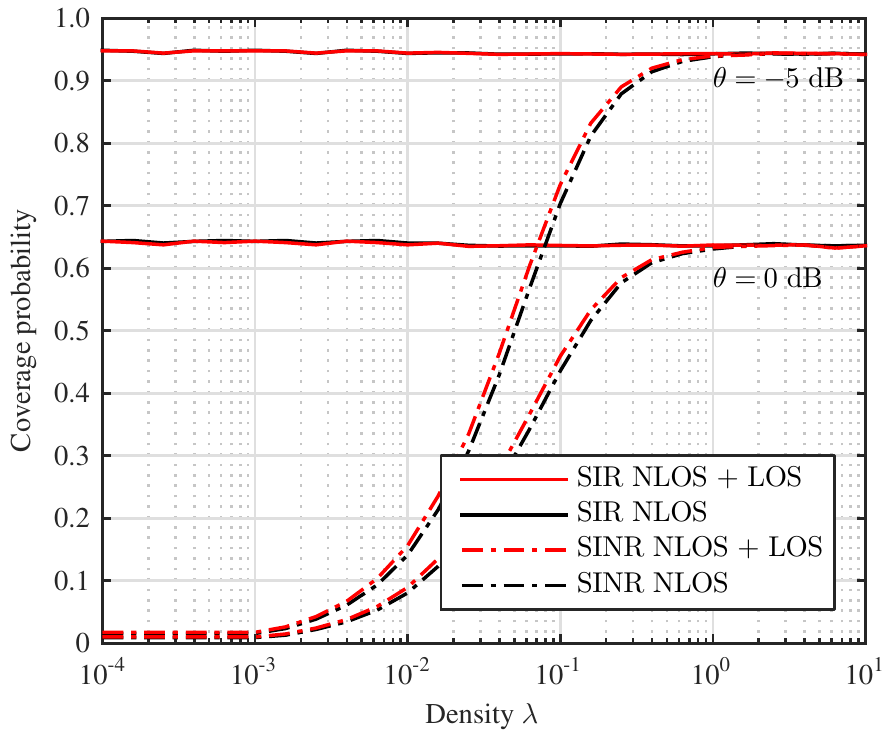}
	\caption{SIR and SINR coverage probability with single-slope path loss ($\alpha=4$): closest (left) and strongest (right) BS association.} \label{fig:Pcov} \vspace{-1mm}
	
\end{figure*}

\section{Simulation results} \label{sec:sim_results}

In this section, we numerically evaluate the derived expressions of the coverage probability and assess the impact of LOS propagation under the assumption of both closest and strongest BS association.

The reported numerical results are obtained by means of Monte Carlo simulations for $10^{5}$ realizations of the PPP with density $\lambda \in [10^{-4},10^{1}]$. 
To begin with, we use the 3GPP LOS probability model in \eqref{eq:3GPP}; then we compare it with the simplified LOS probability model in \eqref{eq:3GPP_simpl} with $D=18$~m. The coverage probabilities in these two settings are compared with the tractable upper bound obtained using the result in Lemma~\ref{lem:closest_simpl}. In the following, we plot both the SIR and the SINR coverage probabilities, where for the latter we assume $\mathrm{SNR}\triangleq 1/\sigma^2=10$~dB; furthermore, we use $\theta=-5$~dB and $\theta=0$~dB as SIR/SINR thresholds. All channels in LOS conditions are characterized by Ricean fading with $K$-factor $K=15$~dB: their amplitude is approximated with a Nakagami-$m$ distribution with $m=(K+1)^{2}/(2K+1)$ rounded to the nearest integer (this corresponds to having $m=17$). We examine both single- and dual-slope path loss models, with $\alpha=4$ for the former and $\alpha_{0}=2.1$, $\alpha_{1}=4$, and $R_{0}=10$~m for the latter.

\begin{figure}
	\centering
	\includegraphics[scale=1]{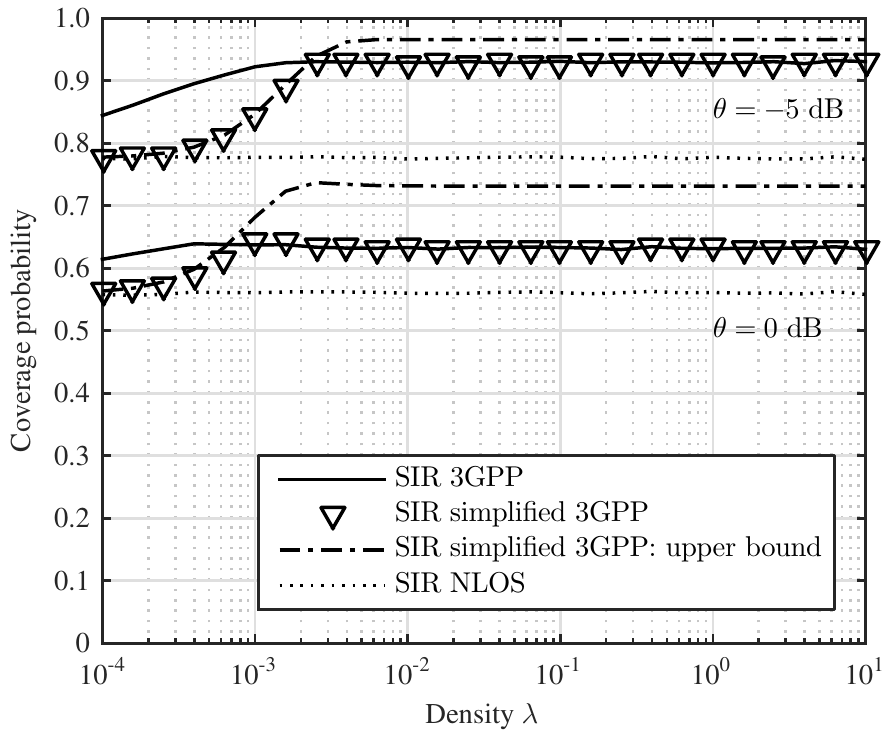}
	\caption{SIR coverage probability with closest BS association and single-slope path loss ($\alpha=4$): 3GPP model, simplified 3GPP model, and tractable upper bound.} \label{fig:ub}
\end{figure}

Figure~\ref{fig:Pcov} plots the coverage probability with the 3GPP model in \eqref{eq:3GPP}. When closest BS association is considered, taking into account the LOS propagation implies a substantial increase of the coverage probability; on the other hand, under strongest BS association, no noticeable difference is observed. This behavior lies in the fact that, with closest BS association, the serving BS is likely to be in LOS conditions, which results in higher SINR as the interferers lie further than the distance from the closest BS; on the contrary, with strongest BS association, the signal from the serving BS is not necessarily in LOS condition and grows at the same rate as the interference. A general and important result is that the impact of LOS propagation with closest BS association is comparable to the effect of switching to strongest BS association in NLOS.

From Figure~\ref{fig:ub}, which focuses only on the SIR coverage probability, it is evident that the simplified 3GPP model resembles very closely the 3GPP model for densities above approximately $\lambda=10^{-3}$. On the other hand, the tractable upper bound on the coverage probability is remarkably close to the simplified 3GPP model in the decreasing part of the curve, whereas it is less accurate in the flat part and becomes tighter for decreasing SIR threshold. Furthermore, we note that the limit in Corollary~\ref{cor:lim} holds already at densities as low as $\lambda=10^{-4}$.

\begin{figure*}[t!]
	\centering
	\includegraphics[scale=1]{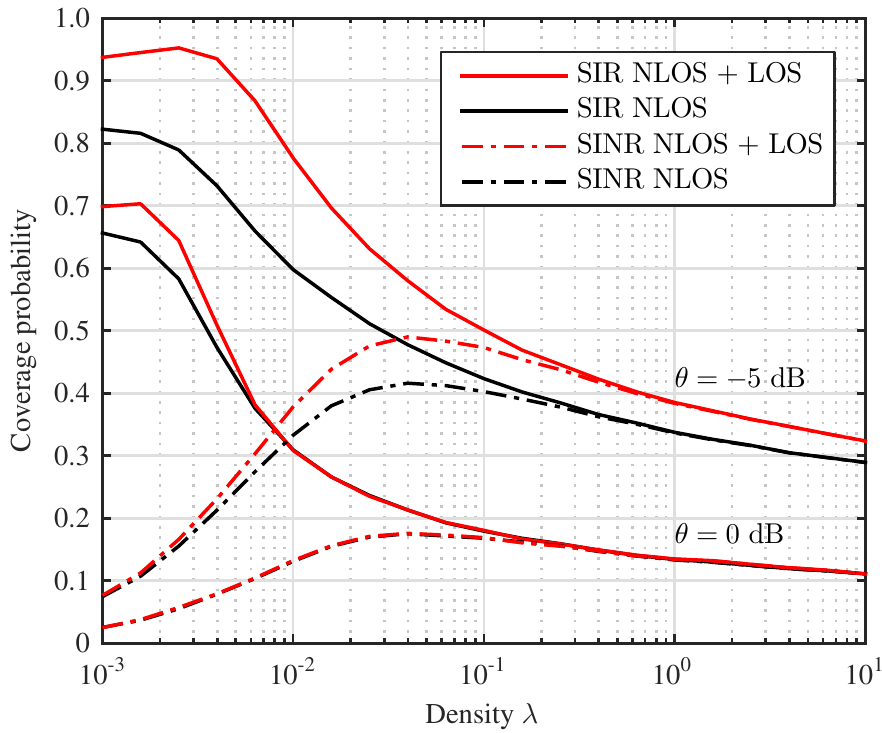}
	\includegraphics[scale=1]{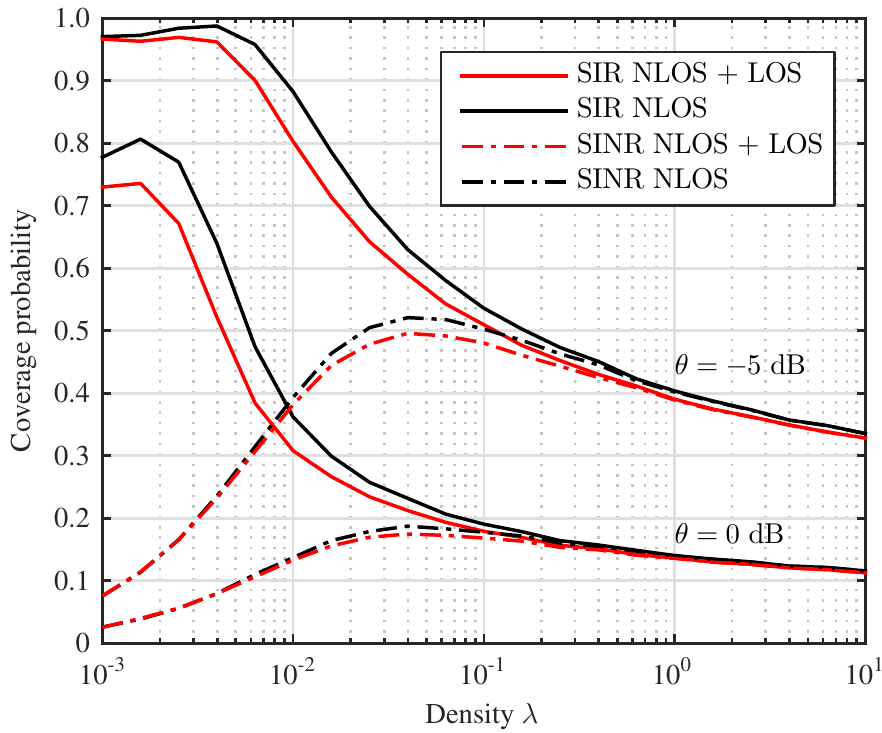}
	\caption{SIR and SINR coverage probability with dual-slope path loss ($\alpha_{0}=2.1$, $\alpha_{1}=4$, and $R_{0}=20$~m): closest (left) and strongest (right) BS association.} \label{fig:Pcov_dual}
\end{figure*}

Lastly, we evaluate the coverage probability using a dual-slope path loss model in Figure~\ref{fig:Pcov_dual}. In this setting, LOS propagation is beneficial for closest BS association and detrimental for strongest BS association. This contrasting behavior can be explained as follows. With closest BS association, the link between the typical UE and the serving BS experiences LOS propagation and path loss exponent $\alpha_{0}$ with high probability in general and with probability one in the ultra-dense regime; on the other hand, the closest interfering BS, which is located further than the nearest neighbor distance, can be either NLOS or LOS conditions depending on the density, and its power is generally lower than the received signal power. Otherwise stated, for low-to-moderate BS densities, the UE is likely to have LOS propagation with the serving BS and NLOS with most of the interfering BS, i.e., the received power is higher than the interference power. With strongest BS association, most of the interfering BS experience LOS and are in the near-field (with path loss exponent $\alpha_{0}$), while the serving BS is not necessarily in the near-field for low-to-moderate densities. Nevertheless, the detrimental effect of LOS on the coverage vanishes in the ultra-dense regime.

\section{Conclusions} \label{sec:conclusions}
In this paper, we propose a stochastic geometry based framework to study the effect of LOS/NLOS propagation on network densification. In particular, we model the LOS propagation as Ricean fading, in contrast to prior work that only assumes Rayleigh fading: this allows for practically relevant performance analysis of UDNs under a more realistic setting. Remarkably, the proposed framework accommodates generalized distance-dependent LOS probability functions; in addition, it encompasses both closest and strongest BS association, as well as single- and multi-slope path loss models. As a particular scenario, we consider the 3GPP LOS probability model and provide a tractable approximation of the coverage probability. Our results provide useful insights into the coverage and throughput performance and show the impact of LOS propagation on BS association. We show that under a standard power-law path loss model, LOS propagation increases the coverage, especially with closest BS association. Moreover, considering dual- or multi-slope path loss models, LOS propagation proves to be beneficial for closest BS association and detrimental for strongest BS association.

\appendices

\section{Proof of Theorem~\ref{th:main}} \label{sec:app_main}

We start by obtaining the expression of the Laplace transform of the interference as

\begin{align}
\setL_{I}^{\los}(s) & = \Exp \bigg[ \exp \bigg( -s \sum_{x \in \Phi} h_{x} r_{x}^{-\alpha} \bigg) \bigg] \\
& = \Exp_{\Phi} \bigg[ \prod_{x \in \Phi} \Exp_{h_{x}} \big[ \exp(- s h_{x} r_{x}^{-\alpha}) \big| r_{x} \big] \bigg].
\end{align}
By sequentially applying Bayes theorem and the PGFL of a PPP, we obtain \eqref{eq:laplace_bayes}--\eqref{eq:laplace_final_step} at the top of the next page, where the operator $\Exp_{h \sim \los}[ \, \cdot \, ]$ denotes the expectation when $h$ is subject to LOS fading, which gives $\Exp_{h \sim \los} \big[ \exp (-s h t^{-\alpha}) \big] = (1 + s t^{-\alpha}/m )^{-m}$, and we note that the corresponding expectation with NLOS conditions is obtained by setting $m=1$. Note that the lower limit of the integral changes according to the type of BS association (i.e., closest or strongest). After substituting this expression into \eqref{eq:laplace_final_step}, we just need to write the coverage probability as a function of the Laplace transform of the interference. We can show that closest BS and strongest BS associations converge to a similar formulation.

\begin{figure*}[t!]
\begin{align}
\label{eq:laplace_bayes} \setL_{I}^{\los}(s) & = \Exp_{\Phi} \bigg[ \prod_{x \in \Phi} \Big( p_{\los}(r_{x}) \Exp_{h_{x} \sim \los} \big[ \exp (- s h_{x} r_{x}^{-\alpha}) \big] + (1 - p_{\los}(r_{x})) \Exp_{h_{x} \sim \nlos} \big[ \exp (-s h_{x} r_{x}^{-\alpha}) \big] \Big) \bigg] \\
\nonumber & = \exp \bigg( -2 \pi \lambda \int_{\nu(r)}^{\infty} \Big( 1 - \Exp_{h \sim \nlos} \big[ \exp (-s h t^{- \alpha} ) \big] \Big) t \, \diff t \bigg) \\
\label{eq:laplace_final_step} & \phantom{= \;} \times \exp \bigg( - 2 \pi \lambda \int_{\nu(r)}^{\infty} p_{\los}(t) \Big( \Exp_{h \sim \nlos} \big[ \exp (-s h t^{-\alpha}) \big] - \Exp_{h \sim \los} \big[ \exp (-s h t^{-\alpha}) \big] \Big) t \, \diff t \bigg)
\end{align}
\hrulefill \vspace{-2mm}
\end{figure*}

\begin{itemize}
\item[1)] With closest BS association and $\theta>1$:
\begin{align} \label{eq:P_C}
\hspace{-3mm} \mathrm{P}_{\mathrm{cov}}^{\mathrm{C}}(\theta) & = \Pr (\sir_{x} > \theta)  = \int_{0}^{\infty} \Pr \big(h > \theta r^{\alpha} I \big| \ r\big) \, \diff F(r)
\end{align}
where $\diff F(r) = 2 \pi \lambda e^{- \pi \lambda r^{2}} \diff r$;
\item[2)] With strongest BS association:
\begin{align}
\mathrm{P}_{\mathrm{cov}}^{\mathrm{S}}(\theta) & = \Pr \bigg( \bigcup_{x \in \Phi} \sir_{x} > \theta \bigg) \\
& = \Exp \bigg[ \sum_{{x} \in \Phi} \mathbbm{1} (\sir_{x}) > \theta \bigg] \\
\label{eq:P_S} & = 2 \pi \lambda \int_{0}^{\infty} \Pr (h > \theta r^{\alpha} I | \ r) r \, \diff r.
\end{align}
\end{itemize}
We note that the only difference in form between \eqref{eq:P_C} and \eqref{eq:P_S} lies in the term $e^{-\pi \lambda r^{2}}$; we can thus unify them by using \eqref{eq:phi}--\eqref{eq:nu} and write

\begin{align}
\mathrm{P}_{\mathrm{cov}}(\theta) = \ & \int_{0}^{\infty} \Pr (h > \theta r^{\alpha} I \big| r) \phi(r) \, \diff r \\
\nonumber = \ & \int_{0}^{\infty} \Big( \big( 1 - p_{\los}(r) \big) \Upsilon_{\nlos}(\theta r^{\alpha}) \\
\label{eq:P_cov} & + p_{\los}(r) \Upsilon_{\los}(\theta r^{\alpha}) \Big) \phi(r)\, \diff r
\end{align}
where we have defined $\Upsilon_{\mathrm{Q}}(z) \triangleq \Exp_{I} \big[ \bar{F}_{\mathrm{Q}} (zI) \big]$, and the sub-index $\mathrm{Q}$ takes the form $\mathrm{Q} = \mathrm{LOS}$ if $h_{x}$ is subject to LOS propagation and $\mathrm{Q} = \mathrm{NLOS}$ otherwise. Thus, we have that
\begin{align}
\label{eq:P_nlos} \hspace{-3mm} \Upsilon_{\nlos}(z) & \triangleq \Exp_{I} \big[ \bar{F}_{\nlos}(z I) \big] = \Exp_{I} \big[ \exp(- z I) \big] = \setL_{I}^{\nlos}(z) \\
\nonumber\Upsilon_{\los}(z) & \triangleq \Exp_{I} \big[ \bar{F}_{\los}(z I) \big] = \Exp_{I} \bigg[ \exp(- m z I) \sum_{k=0}^{m-1} \frac{(m z)^k}{k!} I^{k} \bigg] \\
\label{eq:P_los} & = \bigg[ \sum_{k=0}^{m - 1} \frac{(-s)^{k}}{k!} \frac{\diff^{k}}{\diff s^{k}} \setL_{I}^{\los}(s) \bigg]_{s = m z}.
\end{align}
The proof is finalized by substituting \eqref{eq:P_nlos} and \eqref{eq:P_los} into \eqref{eq:P_cov}.~ \IEEEQED

\section{Proof of Corollary~\ref{cor:lim}} \label{sec:app_lim}

The average distance of the nearest neighbor is given by \cite{Hae05}
\begin{equation}
\Exp \Big[ \min_{x \in \Phi} \{ r_{x} \} \Big] = \frac{1}{2 \sqrt{\lambda}}.
\end{equation}
Therefore, when $\lambda \to 0$, it is not difficult to show that no point of the PPP falls within $r \in [0,D]$. This implies the following:
\begin{align}
& \lim_{\lambda \to 0} \int_{D}^{\infty} \setL_{I}^{\nlos} (\theta r^{\alpha}) \phi(r) \, \diff r = \mathrm{P}_{\mathrm{cov}}^{\nlos}(\theta) \\
& \lim_{\lambda \to 0} \int_{0}^{D} \bigg[ \sum_{k=0}^{m - 1} \frac{(-s)^{k}}{k!} \frac{\diff^{k}}{\diff s^{k}} \widetilde{\setL_{I}^{\los}}(s) \bigg]_{s = m \theta r^{\alpha}} \! \! \! \phi(r) \, \diff r = 0
\end{align}
from which we readily obtain the result in \eqref{eq:lim}. \hfill \IEEEQED

\section{Proof of Lemma~\ref{lem:closest_simpl}} \label{sec:app_closest_simpl}

From $\Upsilon_{\los}(z)$ in \eqref{eq:P_los}, we have
\begin{align}
\Upsilon_{\los}(z) = 1 - \frac{\Exp_{I} \big[ \gamma(m, m z I) \big]}{\Gamma(m)}.
\end{align}
We now use Alzer's inequality \cite{Alz97}
\begin{align}
 \frac{\gamma(m,z)}{\Gamma(m)} > \big( 1 - \exp(- c z) \big)^{m} 
\end{align}
and, since $m>1$, we have $c=(\Gamma(m+1))^{-\frac{1}{m}}$. Now, expanding the expectation term we have
\begin{align}
\hspace{-3mm} \frac{\Exp_{I} \big[ \gamma(m, m z I) \big]}{\Gamma(m)} & \geq \Exp_{I} \big[ \big( 1 - \exp(- c m z I) \big)^{m} \big] \\
& = \Exp_{I} \bigg[ \sum_{k=0}^{m} (-1)^{k} \binom{m}{k} \exp(- c k m z I) \bigg] \\
& = \sum_{k=0}^{m} (-1)^{k} \binom{m}{k} \setL_{I}(ck m z).
\end{align}
Then, observing that $-(-1)^{k} = (-1)^{k+1}$ and that $\binom{m}{0} \setL_{I}(0)=1$, i.e., the term in the sum with $k=0$, the upper bound in \eqref{eq:bounds} readily follows. \hfill \IEEEQED

\vspace{2mm}

\addcontentsline{toc}{chapter}{References}
\bibliographystyle{IEEEtran}
\bibliography{IEEEabrv,ref_Huawei}

\begin{thebibliography}{10}
\providecommand{\url}[1]{#1}
\csname url@samestyle\endcsname
\providecommand{\newblock}{\relax}
\providecommand{\bibinfo}[2]{#2}
\providecommand{\BIBentrySTDinterwordspacing}{\spaceskip=0pt\relax}
\providecommand{\BIBentryALTinterwordstretchfactor}{4}
\providecommand{\BIBentryALTinterwordspacing}{\spaceskip=\fontdimen2\font plus
\BIBentryALTinterwordstretchfactor\fontdimen3\font minus
  \fontdimen4\font\relax}
\providecommand{\BIBforeignlanguage}[2]{{%
\expandafter\ifx\csname l@#1\endcsname\relax
\typeout{** WARNING: IEEEtran.bst: No hyphenation pattern has been}%
\typeout{** loaded for the language `#1'. Using the pattern for}%
\typeout{** the default language instead.}%
\else
\language=\csname l@#1\endcsname
\fi
#2}}
\providecommand{\BIBdecl}{\relax}
\BIBdecl

\bibitem{Bhu14}
N.~Bhushan, J.~Li, D.~Malladi, R.~Gilmore, D.~Brenner, A.~Damnjanovic,
  R.~Sukhavasi, C.~Patel, and S.~Geirhofer, ``Network densification: the
  dominant theme for wireless evolution into {5G},'' \emph{{IEEE} Commun.
  Mag.}, vol.~52, no.~2, pp. 82--89, Feb. 2014.

\bibitem{Que13}
T.~Q.~S. Quek, G.~De~La~Roche, I.~G\"{u}venç, and M.~Kountouris, \emph{Small
  Cell Networks: Deployment, PHY Techniques, and Resource Management}.\hskip
  1em plus 0.5em minus 0.4em\relax New York, NY, USA: Cambridge University
  Press, 2013.

\bibitem{And11}
J.~G. Andrews, F.~Baccelli, and R.~K. Ganti, ``A tractable approach to coverage
  and rate in cellular networks,'' \emph{{IEEE} Trans. Commun.}, vol.~59,
  no.~11, pp. 3122--3134, Nov. 2011.

\bibitem{Dhi12}
H.~S. Dhillon, R.~K. Ganti, F.~Baccelli, and J.~G. Andrews, ``Modeling and
  analysis of {$K$}-tier downlink heterogeneous cellular networks,''
  \emph{{IEEE} J. Sel. Areas Commun.}, vol.~30, no.~3, pp. 550--560, Apr. 2012.

\bibitem{Zha15}
X.~Zhang and J.~G. Andrews, ``Downlink cellular network analysis with
  multi-slope path loss models,'' \emph{{IEEE} Trans. Wireless Commun.},
  vol.~63, no.~5, pp. 1881--1894, May 2015.

\bibitem{Bai14}
T.~Bai, R.~Vaze, and R.~W. Heath, ``Analysis of blockage effects on urban
  cellular networks,'' \emph{{IEEE} Trans. Wireless Commun.}, vol.~13, no.~9,
  pp. 5070--5083, Sep. 2014.

\bibitem{Gal15}
C.~Galiotto, N.~K. Pratas, N.~Marchetti, and L.~Doyle, ``A stochastic geometry
  framework for {LOS}--{NLOS} propagation in dense small cell networks,'' in
  \emph{Proc. {IEEE} Int. Conf. Commun. (ICC)}, London, UK, June 2015.

\bibitem{Din15}
M.~Ding, P.~Wang, D.~Lopez-Perez, G.~Mao, and Z.~Lin, ``Performance impact of
  {LoS} and {NLoS} transmissions in dense cellular networks,'' \emph{{IEEE}
  Trans. Wireless Commun.}, vol.~PP, no.~99, pp. 1--1, 2015.

\bibitem{UMi}
3GPP, ``Technical specification group radio access network; evolved universal
  terrestrial radio access {(E-UTRA)}; further advancements for {E-UTRA}
  physical layer aspects {(Release 9)}. {TR 36.814},'' 2010.

\bibitem{Hae05}
M.~Haenggi, ``On distances in uniformly random networks,'' \emph{{IEEE} Trans.
  Inf. Theory}, vol.~51, no.~10, pp. 3584--3586, Oct. 2005.

\bibitem{Alz97}
H.~Alzer, ``On some inequalities for the incomplete {G}amma function,''
  \emph{AMS Mathematics of Computation}, vol.~66, no. 218, pp. 771--778, Apr.
  1997.

\end{thebibliography}

\end{document}